\documentclass[npg]{copernicus}
\usepackage{amsmath}
\usepackage{bm}
\usepackage{graphicx}
\def\eps{\varepsilon}
\newcommand\Rey{\mbox{\textit{Re}}}

\begin{document}

\title{Accurate estimation of third-order moments from turbulence measurements}

\author[1]{J.~J.~Podesta}
\author[2]{M.~A.~Forman}
\author[1]{C.~W.~Smith}
\author[3]{D.~C.~Elton}
\author[4]{Y.~Mal\'ecot}
\author[5]{Y.~Gagne}

\affil[1]{Space Science Center, University of New Hampshire, Durham, New Hampshire, 03824}
\affil[2]{Department of Physics and Astronomy, State University of New York, Stony Brook, New York, 11794}
\affil[3]{Department of Physics, Applied Physics and Astronomy, Rensselaer Polytechnic Institute, Troy, New York, 12180}
\affil[4]{University Joseph Fourier -- Grenoble I, BP 53, 38041 Grenoble Cedex 9, France}
\affil[5]{Laboratoire des  Ecoulements G\'eophysiques et Industriels, CNRS/UJF/INPG UMR 5519 BP53, 38041 Grenoble, France}

\runningtitle{Accurate estimation of third-order moments from turbulence measurements}

\runningauthor{J.~J.~Podesta}

\correspondence{John Podesta\\ (jpodesta@solar.stanford.edu)}

\received{}
\pubdiscuss{} 
\revised{}
\accepted{}
\published{}


\firstpage{1}

\maketitle

\begin{abstract}
Politano and Pouquet's law, a generalization of Kolmogorov's four-fifths law 
to incompressible MHD, makes it possible to measure the energy cascade rate in 
incompressible MHD turbulence by means of third-order moments.  In hydrodynamics, 
accurate measurement of third-order moments requires large amounts of data because 
the probability distributions of velocity-differences are nearly symmetric and the
third-order moments are relatively small.  Measurements of the energy cascade
rate in solar wind turbulence have recently been performed for the first time, but without 
careful consideration of the accuracy or statistical uncertainty of the required 
third-order moments.  This paper investigates the statistical convergence
of third-order moments as a function of the sample size $N$. 
It is shown that the accuracy of the third-moment $\langle (\delta v_\parallel)^3\rangle$  
depends on the number of correlation lengths spanned by the data set and a method of 
estimating the statistical uncertainty  of the third-moment is developed.  
The technique is illustrated using  both wind tunnel data and solar wind data.  
\end{abstract}

\section{Introduction}

\indent\indent 
In the solar wind, coupling between large- and small-scale turbulence occurs
at kinetic scales defined by the ion gyro-radius and the ion gyro-period.  
At these scales, the turbulent energy cascade undergoes a transition from large 
magnetohydrodynamic (MHD) scales to small plasma kinetic scales where the energy is ultimately 
dissipated by collisionless processes.  Detailed understanding
of the energy cascade process at MHD-scales is a prerequisite for studies of this
coupling.  Here we focus on one particular aspect of MHD-scale turbulence which
is of some practical importance, namely, the determination of the energy
cascade rate from measured data.
\medskip

MHD-scale turbulence in the solar wind is often modeled using the theory of 
incompressible MHD because of its relative simplicity, even though the solar wind is 
known to be compressible.  In the solar wind,
the energy density of MHD turbulence is comparable to the plasma thermal energy
at 1 AU \citep{BelcherDavis:1971} and the turbulent energy cascade is believed to 
significantly heat the solar wind plasma as it flows from $\sim 1$ AU to several tens
of AU.  Theoretical work has shown that plasma heating caused by dissipation of the 
turbulence can likely explain the observed radial temperature profile of the solar wind 
which decreases more slowly than would be the case if the expansion were adiabatic 
\citep{Matthaeus_Zank:1996,Zank:1999,Matthaeus:1999,Smith:2001,Isenberg:2003}.  
To refine these theories, accurate measurements of the energy cascade rate are needed.
Recently, the energy cascade rate $\eps$ has been directly measured for the first
time in the solar wind using a generalization of Kolmogorov's four-fifths law
\citep{MacBride:2005,Sorriso-Valvo:2007, MacBride:2008,Marino:2008}.  
Before discussing this, it may be helpful to provide some background information on 
Kolmogorov's four-fifths law.
\medskip

For turbulent flows in ordinary incompressible fluids such as air or water
the energy cascade rate $\epsilon$ is often measured {\it indirectly} by
means of the energy dissipation rate
\begin{equation}
\eps_{\mbox{\scriptsize diss}} = 15\nu \bigg\langle \bigg(\frac{\partial v}{\partial x}
\bigg)^{\!\! 2} \bigg\rangle, 
\label{diss}
\end{equation}
where $\nu$ is the kinematic viscosity and the coefficient 15 arises from the
assumption that the turbulence is isotropic \citep[p.~134]{Pope:2000}. The energy cascade rate 
can also be measured {\it directly} by means of Kolmogorov's four-fifths law
\begin{equation}
\big \langle (\delta v_\parallel)^3 \big\rangle =-{\textstyle \frac{4}{5}} \epsilon r,
\label{K45}
\end{equation}
valid for isotropic turbulence, where 
\begin{equation}
\delta v_\parallel(r) = \big[ \bm v(\bm x+\bm r)-\bm v(\bm x)\big] \cdot \hat{\bm e}_r
\end{equation}
is the component of the velocity fluctuation in the direction of the displacement 
$\bm r$ and the lengthscale $r$ lies in the inertial range \citep{Kolmogorov:1941, Frisch:1995}. 
Note that Kolmogorov's four-fifths law (\ref{K45}) is independent of the kinematic viscosity
$\nu$ and can be applied even when the kinematic viscosity is unknown, but
the accurate evaluation of the third-order moment (\ref{K45}) requires
much more data than the second-order moment (\ref{diss}).
\medskip

Kolmogorov's four-fifths law was originally derived for homogeneous isotropic 
turbulence and a similar law was later derived by Monin for homogeneous 
{\it anisotropic} turbulence; see \citet{Podesta_Forman:2007} for references.  
\citet{Politano:1998a,Politano:1998b} generalized these 
fundamental results of Kolmogorov and Monin from the theory of incompressible 
hydrodynamic turbulence to incompressible MHD turbulence. 
It is important to emphasize that Politano and Pouquet's law holds for both
isotropic and anisotropic turbulence, although this fact was 
not explicitly mentioned by \citet{Politano:1998a}. This is especially important
in MHD where statistical isotropy may not hold in the presence 
of an ambient magnetic field.  A derivation of Politano and Pouquet's law
which is similar to Frisch's derivation of Kolmogorov's four-fifths law is
given by \citet{Podesta:2008}.
\medskip

Politano and Pouquet's law has recently been applied to obtain direct
measurements of the energy cascade rate in the solar wind under the 
simplifying assumption that the turbulence is isotropic 
\citep{MacBride:2005,Sorriso-Valvo:2007, MacBride:2008,Marino:2008}.
\citet{MacBride:2008} have also investigated a non-isotropic 1D/2D hybrid model that
is believed to be descriptive of the solar wind.  The method used in all these studies 
consists of the evaluation of certain third-order moments which are
similar to those in equation (\ref{K45}), except that for incompressible MHD 
turbulence the relevant third-order moments contain combinations of velocity and 
magnetic field fluctuations (or, equivalently, fluctuations in the Elsasser variables).
From the linear scaling of these third-order moments, the energy cascade rate 
is obtained without any knowledge of
the dissipation processes or the viscous and resistive dissipation
coefficients in the the solar wind.  
\medskip

The solar wind studies mentioned above have not given careful consideration
to the convergence properties of third-order moments which raises the question:
how much data is required to accurately estimate the third-order moments?  
The study by Sorriso-Valvo et al. (2007) used approximately 2000 data points
to compute the third-order moments while the study by MacBride et al. (2008)
used close to $10^6$ data points.  The purpose of the present work 
is to investigate the accuracy of third-order moments as a function of the
sample size $N$ (the number of data points used in the analysis).  An important conclusion
is that the accuracy of third-order moments depends on the number of correlation lengths
spanned by the data set (defined below).  The number of correlation lengths
determines the accuracy and statistical uncertainty of third-order moments computed from  
measured data, not the number of data points $N$.  It turns out that for
turbulence studies where the skewness of the distribution is
usually small the accurate estimation of third-order moments requires large amounts of
data.  The reason is partly because
the third moment is not an absolute moment $\langle |x|^3 \rangle$ but a signed 
moment $\langle x^3 \rangle$ and, therefore, is subject to cancellation effects.
The theory describing the convergenge of these moments is
illustrated using turbulence data from the ONERA/Modane wind tunnel.  The same techniques
can be applied to third-order moments in
solar wind turbulence which exhibit similar behavior.


\medskip


\section{Theory}

\subsection{Uncorrelated time series}

\indent\indent 
Given $N$ independent samples $x_1, x_2, \ldots, x_N$
drawn randomly from a probability distribution $f(x)$, the moments of the
distribution can be estimated as
\begin{equation} 
\langle x \rangle \simeq  \frac{1}{N} \sum_{n=1}^N x_n,     
\end{equation}
\begin{equation} 
\langle x^2 \rangle \simeq  \frac{1}{N} \sum_{n=1}^N x_n^2, 
\end{equation}
\begin{equation} 
\langle x^3 \rangle \simeq  \frac{1}{N} \sum_{n=1}^N x_n^3, 
\end{equation}
etc.  Now focus attention on the third moment and let
\begin{equation}
M_3(N)=\frac{1}{N} \sum_{n=1}^N x_n^3. \label{M3}
\end{equation}
Note that $M_3(N)$ is itself  a random variable whose probability distribution can 
be derived, in principle, from the pdf $f(x)$ of the random variable $x$.  
Now suppose that we know the mean and standard deviation of the random variable 
$M_3(N)$ denoted by $\mu_3$ and $\sigma_3$, respectively.  If $\langle x^3 \rangle \ne 0$, 
then for $M_3(N)$ to be an accurate estimate of $\langle x^3 \rangle$ the
standard deviation must be small compared to the mean, that is, 
\begin{equation}
\left|\frac{\sigma_3(N)}{\mu_3(N)}\right| \ll 1. 
\label{ineq}
\end{equation}
This condition can be used to estimate the sample size $N$ required to obtain 
an accurate estimate of the third moment.  Hereafter, it is assumed that
$\langle x^3 \rangle \ne 0$.
\medskip

From equation (\ref{M3}) and the independence of the $x_n$, the first and
second moments  of $M_3(N)$ are 
\begin{equation}
\langle M_3(N) \rangle =\frac{1}{N} \sum_{n=1}^N \langle x_n^3 \rangle = \langle x^3 \rangle
\label{mean}
\end{equation}
and
\begin{multline}
\big\langle \,[M_3(N)]^2 \big\rangle =\frac{1}{N^2} \sum_{n=1}^N \sum_{m=1}^N 
\langle x_n^3 x_m^3 \rangle \\ = \frac{1}{N} \langle x^6 \rangle + \frac{N-1}{N}\langle x^3 \rangle^2.
\end{multline}
Thus, the variance is
\begin{equation}
\sigma_3^2(N) = \big\langle \,[M_3(N)]^2 \big\rangle - \langle M_3(N) \rangle^2 
=\frac{1}{N} \big[ \langle x^6 \rangle -\langle x^3 \rangle^2 \big]
\label{var}
\end{equation}
and
\begin{equation}
\left|\frac{\sigma_3}{\mu_3}\right| =\frac{1}{\sqrt N} 
\left| \frac{\langle x^6 \rangle}{\langle x^3 \rangle^2}-1 \right|^{1/2}. 
\label{R}
\end{equation}
The value of $N$ required to make the last equation small $(\ll 1)$ depends on the ratio 
$\langle x^6 \rangle/\langle x^3 \rangle^2$ and, therefore, depends on the distribution
function $f(x)$.  If the ratio $\langle x^6 \rangle/\langle x^3 \rangle^2$ is on the order of
unity, then $N\gtrsim 10^3$ may be adequate.  But, if this ratio
is much larger than unity, then $N$ will have to increase accordingly.
The relation (\ref{R}) shows that to increase the accuracy of the third-moment
by a factor of ten requires an increase in the sample size $N$ by a factor
of 100.  This slow rate of convergence imposes practical
limitations on estimates of third-order moments obtained from experimental data.
\medskip

\subsection{Correlated time series}

\indent\indent 
For applications to turbulence, the random variable $x$ is a velocity difference
such as $\delta v_\parallel$ and the sequence $x_1, x_2, \ldots, x_N$
is usually not mutually stochastically independent.  For example, two velocity increments
that overlap in space or time are usually correlated to some degree.  
In this case, the number of ``independent
samples'' $N$ in the above theory should be replaced by the number of correlation
lengths of the quantity under consideration.  For the third-order moment
$M_3$ it is necessary to use the correlation length or correlation time $\tau_c$ of the 
time series $y_n=x_n^3$ and replace $N$ in equation (12) by the number of correlation
lengths
\begin{equation}
N_c = \frac{T}{\tau_c},
\label{N_c}
\end{equation}
where $T=Nt_s$ is the temporal record length and $t_s$ is the sampling time.  
Thus, equation (12) takes the modified form
\begin{equation}
\left|\frac{\sigma_3}{\mu_3}\right| =\frac{1}{\sqrt N_c} 
\left| \frac{\langle x^6 \rangle}{\langle x^3 \rangle^2}-1 \right|^{1/2} 
\end{equation}
or, equivalently,
\begin{equation}
\left|\frac{\sigma_3}{\mu_3}\right| =
\bigg(\frac{n}{N}\bigg)^{1/2} 
\left| \frac{\langle x^6 \rangle}{\langle x^3 \rangle^2}-1 \right|^{1/2}, 
\label{Rnew}
\end{equation}
where $\tau_c=nt_s$.  Note that this is almost the same as equation (12)
except for an additional scale factor $n^{1/2}$.  Because the
correlation times of the sequences $x_n$ and $y_n=x_n^3$ can be different
it is important to use the correlation time $\tau_c$ of the sequence $y_n=x_n^3$
in equations (\ref{N_c})--(\ref{Rnew}) when analyzing the third-order moment.
\medskip

\section{Textbook example}

\indent\indent 
An example is now given to illustrate the theory described in section 2.  
Consider the slightly skewed distribution function
\begin{equation}
f_0(x) = \frac{1}{\sqrt{2\pi}} [1+\alpha (x+\alpha)] e^{-(x+\alpha)^2/2},
\label{df}
\end{equation}
where the parameter $\alpha$ characterizes the skewness
of the distribution.  The distribution has zero mean, $\langle x \rangle = 0$, and reduces to
the Gaussian distribution when $\alpha=0$.  If $y=x+\alpha$, the pdf of $y$
is
\begin{equation}
f(y) = \frac{1}{\sqrt{2\pi}} (1+\alpha y) e^{-y^2/2}.
\label{pdf_y}
\end{equation}
The moments of the distribution function
(\ref{pdf_y}) can be computed from the characteristic function
\begin{equation}
F(k) = \int_{-\infty}^{\infty} f(y) e^{iky} \, dy = (1+i\alpha k) e^{-k^2/2}
\label{cf}
\end{equation}
by means of the well known relations
\begin{equation} \begin{array}{l}
F'(0) = i \langle y \rangle,     \\
F^{\prime\prime} (0) = i^2 \langle y^2 \rangle, \\
F^{\prime\prime\prime}(0) = i^3 \langle y^3 \rangle, \end{array}
\end{equation}
etc.  After some tedious calculations, the first six moments are found to be
\begin{equation} \begin{array}{l}
\langle y \rangle = \alpha,     \\
\langle y^2 \rangle  = 1,       \\
\langle y^3 \rangle  = 3\alpha, \\
\langle y^4 \rangle = 3,     \\
\langle y^5 \rangle  = 15\alpha,       \\
\langle y^6 \rangle  = 15. \end{array}
\end{equation}
Thus, from the relation $\langle x^n \rangle = \langle (y-\alpha)^n \rangle$,
the first six moments of the variable $x$ are
\begin{equation} \begin{array}{l}
\langle x \rangle = 0,     \\
\langle x^2 \rangle  = 1-\alpha^2,       \\
\langle x^3 \rangle  = 2\alpha^3, \\
\langle x^4 \rangle = 3-6\alpha^2-3\alpha^4,     \\
\langle x^5 \rangle  = 20\alpha^3+4\alpha^5,       \\
\langle x^6 \rangle  = 15-45\alpha^2-45\alpha^4-5\alpha^6. \end{array}
\end{equation}
For the particular distribution function (\ref{df}), the ratio (\ref{R}) 
takes the form
\begin{equation}
\left|\frac{\sigma_3}{\mu_3}\right| =\frac{1}{\sqrt N} 
\left| \frac{15-45\alpha^2-45\alpha^4-5\alpha^6}{4\alpha^6}-1 \right|^{1/2}. 
\label{Rx}
\end{equation}
For $\alpha = 0.25$, for example, this becomes
\begin{equation}
\left|\frac{\sigma_3}{\mu_3}\right| =\frac{111}{\sqrt N}.
\label{Rxx}
\end{equation}
This ratio is small if $N\gtrsim 10^6$.  In this idealized
example where the distribution function $f(x)$ is known, the number of samples
required to obtain an accurate estimate of the third moment from
experimental data can be computed explicitly.  For turbulence data
acquired in the laboratory, such precise estimates cannot be computed
{\it a priori} because the distribution function $f(x)$ is unknown.
\medskip


\section{A practical approach}

\indent\indent 
When working with experimental turbulence data the distribution function $f(x)$ 
is usually unknown so that the ratio $\langle x^6 \rangle/\langle x^3 \rangle^2$
in equation (\ref{R}) cannot be evaluated.  
A practical approach is to compute the third-moment $M_3(N)$ from the data 
and then construct the empirical distribution function for $M_3(N)$, where $N$ is
now fixed (a constant).  This too may be impractical because of the large number 
of data points required.  However, the distribution function for $M_3(N)$ contains
more information than is needed.  Just a few independent estimates of the third-moment $M_3(N)$, 
perhaps 10, may be sufficient to obtain a rough estimate of the ratio in equation
(\ref{ineq}).  The number of samples $N$ can then be increased until the ratio so 
obtained satisfies the inequality (\ref{ineq}).  This is a simple way of 
controlling the accuracy of third-order moments estimated from turbulence 
measurements.  The effectiveness of the method can be improved by increasing the 
number of independent estimates of $M_3(N)$ used to compute the mean and
standard deviation. The standard deviation obtained from the data provides a rough 
estimate of the 1-$\sigma$ error for the third-order moment.
\medskip

A more precise analysis can be performed by computing histograms, means, and
standard deviations of the third-moment $M_3(N)$ for progressively larger values of $N$.  
Accurate values of the mean $\mu_3(N)$ and standard deviation $\sigma_3(N)$ can be 
obtained for values of $N$ much smaller than the record size.  Fitting the measured ratio 
$|\sigma_3/\mu_3|$  to the functional form $A/\sqrt{N}$, where $A$
is an adjustable parameter, it is then possible 
to extrapolate the ratio $|\sigma_3/\mu_3|$ to larger $N$ where direct calculations
have poor statistics or are unattainable as a consequence of the limited record size.
An alternate fitting function is $A/N^p$ where $A$ and $p$ are two adjustable
parameters.
From this extrapolation it is possible to determine the value of the sample size $N$
required to achieve any desired accuracy of the ratio $|\sigma_3/\mu_3|$ and, therefore,
of the third-order moment $M_3(N)$.  This approach is accurate and effective 
as long as sufficient data are available and requires no {\it apriori} knowledge 
of the distribution function $f(x)$ or its moments.  The same technique can also be applied to
accurately determine moments of any order provided sufficient data are available.
\medskip

\section{Illustration using wind tunnel data}

\indent\indent 
The technique described in the previous section shall now be applied to study 
turbulence data from the ONERA wind tunnel in Modane, France, characterized by a 
Taylor-scale Reynolds number $R_\lambda \simeq 2500$ 
\citep{Kahalerras:1998, Malecot:2000, Gagne:2004}.  
This particular data set is a time series consisting of $1.44 \times 10^7$
data points with a sampling rate of 25 kHz and an average velocity of 20.37 m/s.  
The inertial range extends from $\sim \!10$ Hz to $\sim \!10^3$ Hz as inferred
from the power spectrum shown in Figure 1.  
\begin{figure}[t]
\vspace*{2mm}
\begin{center}
\includegraphics[width=8cm]{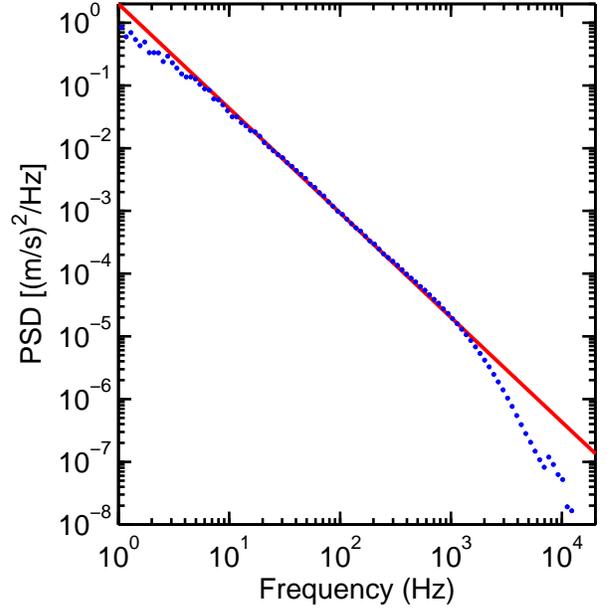}
\end{center}
\caption{Power spectrum of longitudinal velocity fluctuations measured in the
Modane wind tunnel (blue dots).  For comparison, the red line is proportional
to $f^{-5/3}$, Kolmogorov's theoretical inertial range spectrum.  
The inertial range extends from approximately $10$ Hz to $10^3$ Hz.  The onset of the 
dissipation range is indicated by the change in slope around $10^3$ Hz.}
\end{figure}
Now, consider the third-order moment
\begin{equation}
\langle (\delta v_\parallel)^3 \rangle =\big\langle [v(t)-v(t+\tau)]^3 \big\rangle
\end{equation}
where $\tau=1/50$ s or, equivalently, $f=1/\tau= 50$ Hz.  
This time lag is chosen for study because it lies inside the inertial range 
displayed in Figure 1.  
\medskip

The third-order moment is computed using a contiguous series of N data points.
A set of N contiguous data points is called a data block.  A series of successive
data blocks are then used to compute a series of third-order moments, one for each data
block.  The first data point in a given data block is separated from the 
first point of the next successive data block by an offset $m$ where, ideally,
$m=N$.  When the sample size $N$ is not small compared to
the record size, smaller values of the offset $m$ are used so that the total
number of data blocks yields a sufficient statistical sample. Note,
however, that when the offset $m$ is smaller than $N$ the third-order moments
obtained from successive data blocks become dependent (because the blocks overlap)
and, consequently, to obtain
good statistics it is advisable not to let $m$ become much smaller than $N$.  
This tradeoff is unavoidable when working with records of finite length.
\medskip

The set of third-order moments obtained for a given sample size $N$ are used to generate a 
histogram of third-order moments as shown in Figure 2.  
\begin{figure*}[t]
\vspace*{2mm}
\begin{center}
\includegraphics[width=12cm]{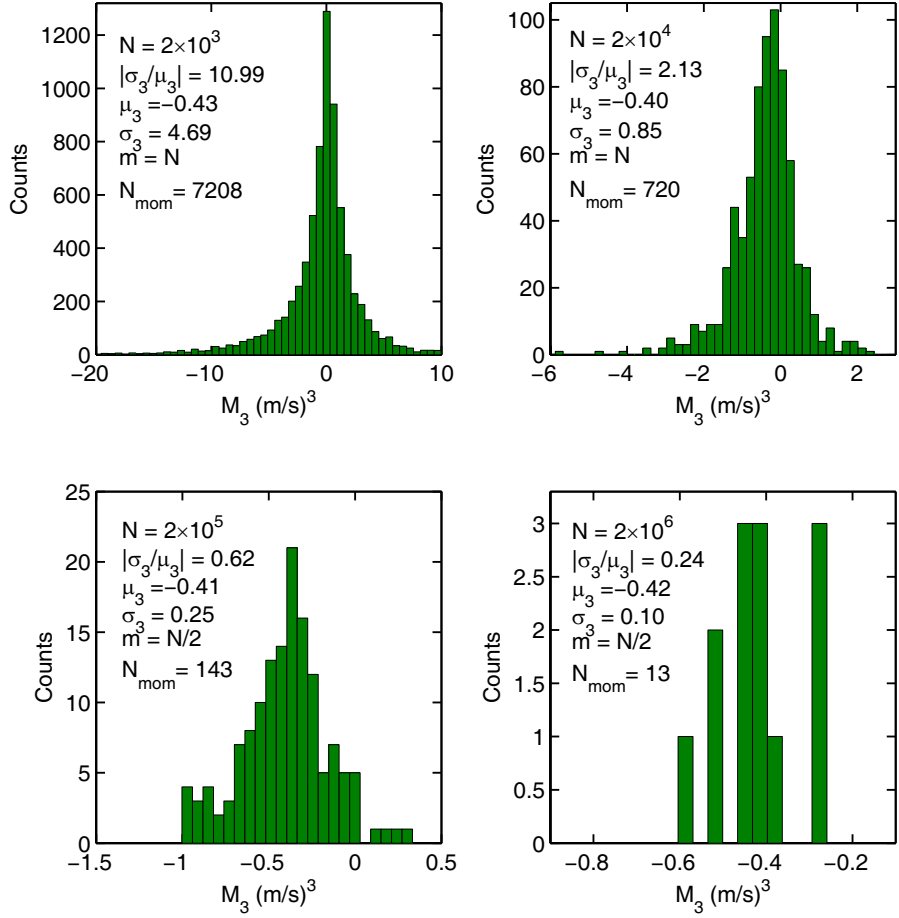}
\end{center}
\caption{Histograms of the third-order moments $M_3(N)$ for the time lag $\tau=20$ ms
obtained from Modane 
wind tunnel data using samples of size $N=2\times 10^3$, $N=2\times 10^4$, $N=2\times 10^5$, 
and $N=2\times 10^6$. In each case, the total number of moments computed is $N_{\rm mom}$,
the offset between adjacent data blocks is $m$, and the sample mean and standard deviation
$\mu_3$ and $\sigma_3$ have units (m/s)$^3$.  For the case $N=2\times 10^3$ the range of 
$M_3$ values in the histogram extends from $-70.23$ to 49.80 (not shown).  
The range of $M_3$ values shown in each of the other three histograms are all inclusive.}
\end{figure*}
The number of third-order moments $N_{\rm mom}$ 
is equal to the number of data blocks employed in the calculation.
As expected, the width of the distributions as measured by the standard deviation
is a decreasing function of $N$.  Moreover, the results for the ratio of the
standard deviation to the mean are in approximate agreement with 
the $N^{-1/2}$ convergence rate predicted by the theory in section 2.  Thus, increasing 
$N$ by a factor of ten causes a decrease in the ratio of the
standard deviation to the mean by a factor of $\sim 3$.  
\medskip

The third-order moment obtained using all available data is $M_3=-0.406$ (m/s)$^2$.
Note that the mean $\mu_3$ displayed in Figure 2 is approximately independent
of $N$.  This is to be expected because for any sequence of numbers partioned
into successive non-overlapping blocks, the average of the mean values for 
each data block is equal to the mean value of the entire record.  In practice,
the data blocks may not completely cover the given record because the
record size is not divisible by $N$ and, therefore, the equality is only
approximate.  This explains the approximate independence of $\mu_3(N)$
versus $N$ in Figure 2.  See also equations (\ref{mean}) and (\ref{var})
which predict that $\mu_3(N)$ is independent of $N$ and $\sigma_3(N)\sim 1/N^{1/2}$.
\medskip

How much data is required to obtain an accurate estimate of the third-order moment?  
This depends, of course, on the level of error which is tolerable for the application 
at hand.  The relative error is measured by the ratio $|\sigma_3/\mu_3|$.
This quantity is plotted as a function of $N$ in Figure 3 (upper plot).  
\begin{figure}[t]
\vspace*{2mm}
\begin{center}
\includegraphics[width=7.5cm]{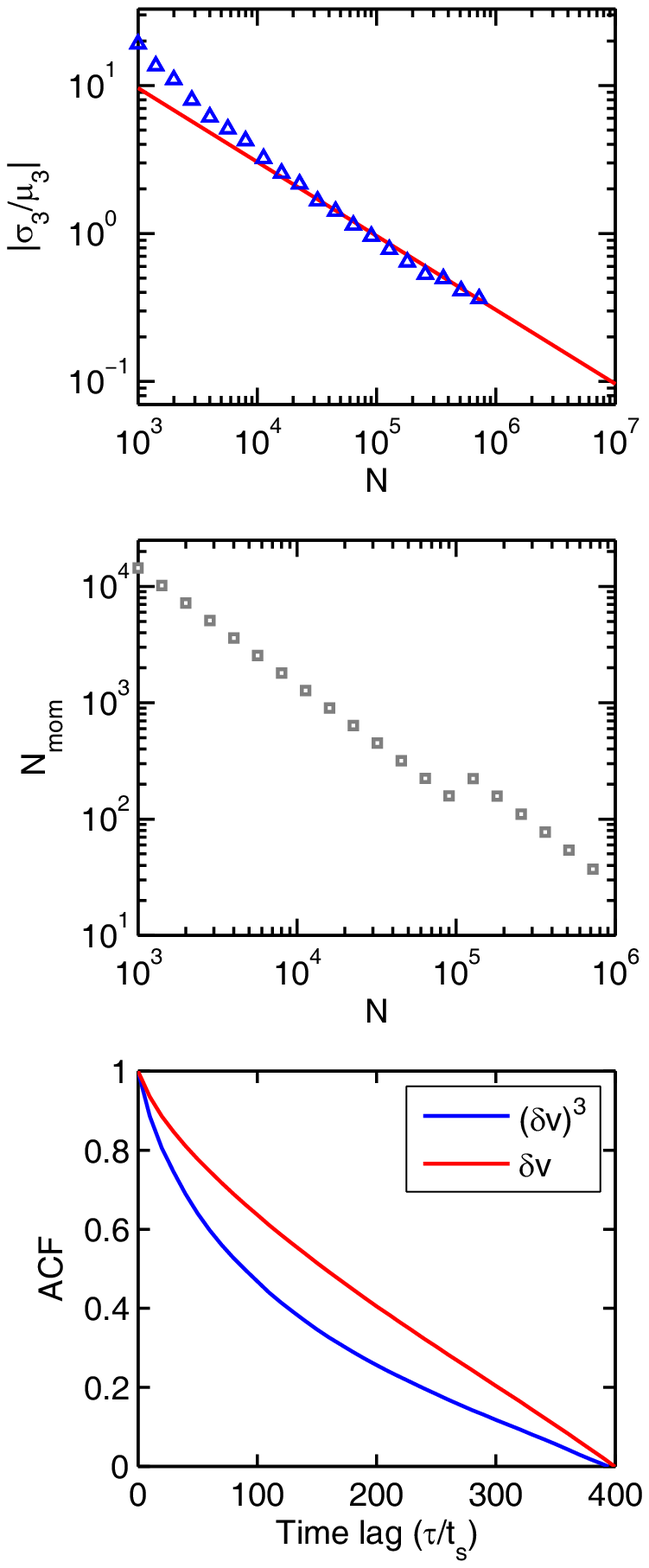}
\end{center}
\caption{Empirical results for the quantity $|\sigma_3/\mu_3|$ as a function of the 
sample size $N$ for the time lag $\tau=20$ ms obtained using Modane wind tunnel data 
(upper plot); the theoretical curve $304/\sqrt{N}$, equation (15), is drawn in red.
The number $N_{\rm mom}$ of third-moments $M_3(N)$ 
used to compute the mean $\mu_3$ and standard deviation $\sigma_3$ are shown in the middle plot.
The autocorrelation function of the difference series $(\delta v_n)^3$ is used to
determine the correlation time $\tau_c\simeq 90 t_s$ used in equation (15);
$\tau_c$ is the point where ACF $=0.5$ (bottom plot).}
\end{figure}
To ensure a reasonably large number of third-moments
$N_{\rm mom}$, the offset $m$ between successive data blocks is $m=N$ when
$N<10^5$ and $m=N/2$ when $N>10^5$.  The number of third-moments $N_{\rm mom}$ is 
shown in the middle plot in Figure 3.  The theoretical relation (15)
takes the form 
\begin{equation}
\left| \frac{\sigma_3(N)}{\mu_3(N)}\right| \simeq \frac{304}{\sqrt N},
\label{fit}
\end{equation}
where the value 304 is obtained using the empirical values of the
sixth moment $M_6=169.5$ (m/s)$^6$, the third moment $M_3=-0.4062$ (m/s)$^6$, 
and the correlation time $\tau_c\simeq 90 t_s$
defined as the time where the autocorrelation function equals 1/2 (Figure 3).
Inspection of the theoretical curve,
the red line in Figure 3, shows that to achieve the level of precision 
$|\sigma_3/\mu_3|\le 0.1$ would require $N\gtrsim 10^7$ data points or, equivalently
$N_c\gtrsim 1.1\times 10^5$ correlation lengths.  
The total number of data points contained in the data set is $1.44\times 10^7$.
\medskip

One can see from this example that accurate estimation of third-order
moments from turbulence data requires a very large record length.
Under circumstances where sufficiently large data sets are not available, the techniques
described here and in the last section can be used to estimate the errors
in the third-moment as quantified by the standard deviation $\sigma_3$
and the empirical ratio $|\sigma_3/\mu_3|$.  
\medskip

So far in this section the analysis of the third-moment has been carried out for 
one time lag $\tau=1/50$ s.  The same analysis can be carried out for many
different time lags and, in this case, the error $|\sigma_3/\mu_3|$ is typically an increasing
function of time lag $\tau$ throughout the inertial range (for a fixed sample size $N$).
The third-order moment as a function of the time lag $\tau$ computed using
all available data is shown in the upper plot in Figure 4. For the data shown in Figure 4,
\begin{figure}[t]
\vspace*{2mm}
\begin{center}
\includegraphics[width=7cm]{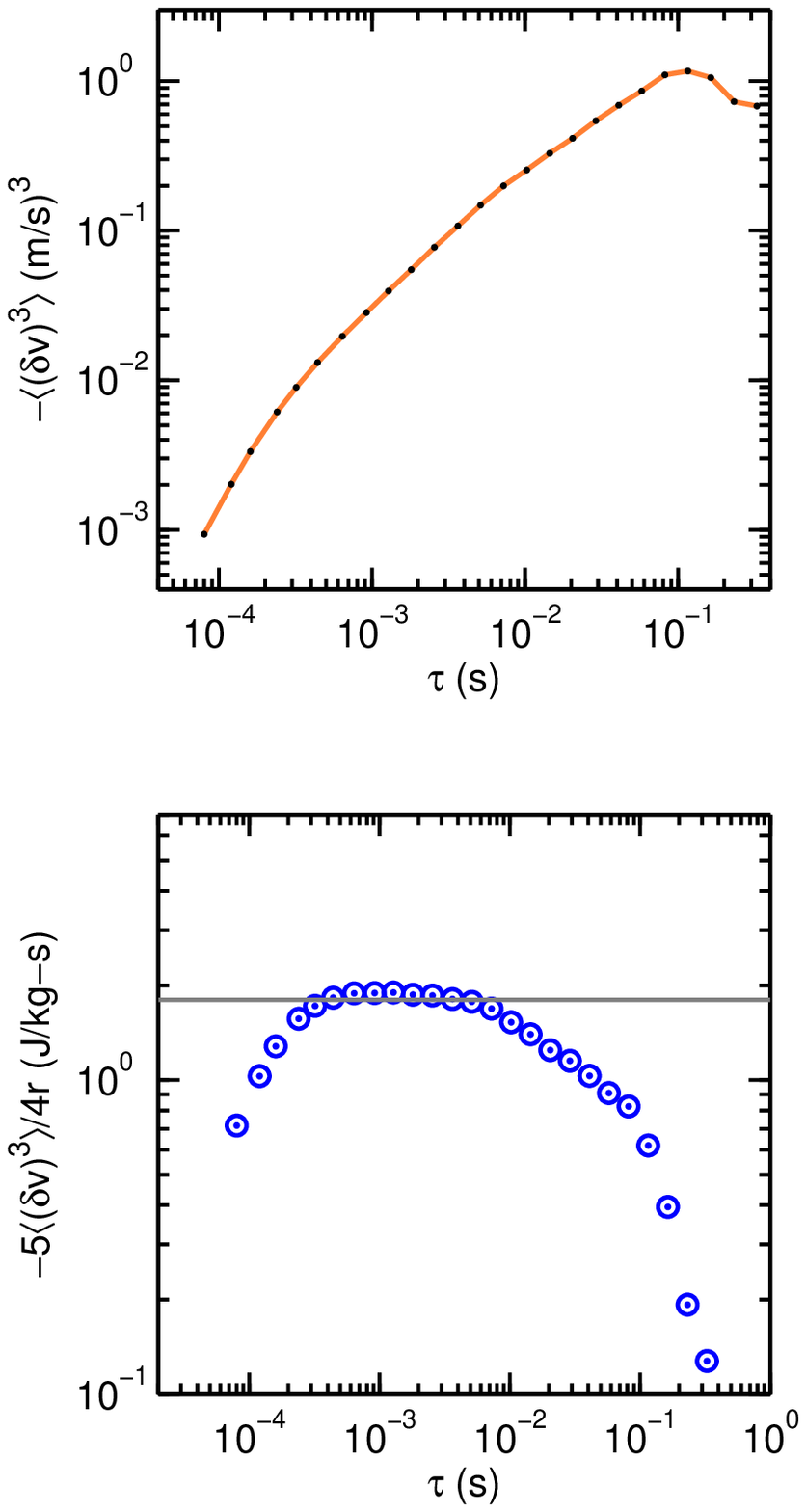}
\end{center}
\caption{The third-order moment $\langle (\delta v_\parallel)^3 \rangle$
versus time lag $\tau$ computed using the entire record of Modane wind tunnel data
(upper plot). The ratio $-5\langle (\delta v_\parallel)^3 \rangle/4r$ versus
$\tau$, where $r=\bar v \tau$ is the spatial separation (lower plot).  The horizontal
line in the lower plot is the value 1.8 J/kg-s.  The data for $\tau \gtrsim 0.1$ s is
uncertain and should be disregarded.}
\end{figure}
estimates show that the relative error $|\sigma_3/\mu_3|$ lies approximately in the range 
$0.09<|\sigma_3/\mu_3| \lesssim 0.3$ for $\tau<0.1$ s and $|\sigma_3/\mu_3| \gtrsim 0.3$ 
for $\tau>0.1$ s.  Hence, the third-moments are sufficiently accurate for the 
present purpose only for $\tau<0.1$ s.  
\medskip

To estimate the energy cascade rate
$\eps$ using Kolmogorov's four-fifths law (\ref{K45}), the quantity 
$-5\langle (\delta v_\parallel)^3 \rangle/4r$ is plotted versus $\tau$ in the 
lower panel in Figure 4.  
Note that compressibility effects are negligible because the Mach number is much less than 
unity and, therefore, the application of Kolmogorov's four-fifths law is justified.  In the 
lower panel in Figure 4, the data lie approximately on the horizontal 
line $\eps = 1.8$ J/kg-s throughout the range $2 \times 10^{-4}<\tau<10^{-2}$ s.  
Thus, the value of the energy cascade rate determined by Kolmogorov's
four-fifths law is $\eps \simeq 1.8$ J/kg-s.  Note that the range 
$2 \times 10^{-4}<\tau<10^{-2}$ s where where an apparent plateau is formed does 
not coincide with the inertial range $10^{-3}< \tau<10^{-1}$ s inferred from Figure 1.
Because the dissipation range lies just beyond the spectral
break near $10^3$ Hz in Figure 1 (Pope, page 237), this implies that the region
where the four-fifths law holds includes part of the dissipation range.
It is also puzzling why the four-fifths law breaks down for $\tau \gtrsim 10^{-2}$ s
in Figure 4 since the inertial range appears to extend to $\tau \simeq 10^{-1}$ s
in Figure 1.  Consequently, Kolmogorov's four-fifths law does not hold throughout the
entire inertial range as the theory seems to predict.  
The reason for these discrepancies is unknown at the moment.  However, 
results for the scaling of the third-order moment in Figure 4 are in agreement with
\citet{Gagne:2004} who analyzed the same Modane data.
\medskip

An independent estimate of the energy cascade rate
is obtained using equation (\ref{diss}). If the measured signal is approximated
by the truncated Fourier series
\begin{equation}
v(t) = \sum_{k=-N/2}^{(N/2)-1} V_k \exp\bigg(\frac{i2\pi kt}{T}\bigg),
\end{equation}
where $N$ is the record length, $T=N\Delta t$, and $\Delta t$ is the sampling time,
then the time average of $(\partial v/\partial t)^2$ is given by
\begin{equation}
\bigg\langle \bigg(\frac{\partial v}{\partial t}
\bigg)^{\!\! 2} \bigg\rangle = \bigg(\frac{2\pi}{T} \bigg)^{\!\! 2} \sum_{k=-(N/2)+1}^{N/2}
|kV_k|^2, 
\label{deriv}
\end{equation}
where
\begin{equation}
V_k = \frac{1}{N} \sum_{n=0}^{N-1} v_n \exp\bigg(\frac{i2\pi kn}{N}\bigg)
\end{equation}
is the discrete Fourier transform of the sequence $v_n=v(n\Delta t)$
which is easily evaluated using the FFT.  Using the entire data record
to evaluate (\ref{deriv}) and the value $\nu=2\times 10^{-5}$ m$^2$/s, 
the energy dissipation rate obtained from equation (\ref{diss}) is $\eps_{\rm diss} = 2.6$ J/kg-s.
\medskip

The spatial separation between two consecutive measurements $\ell =\bar v \Delta t =0.8$
mm is roughly three times the Kolmogorov scale $\eta\simeq 0.3$ mm; the
normalized wavenumber is $k\eta \simeq 2.4$.  Because most of the dissipation occurs  
in the wavenumber range $k\eta \lesssim 1$ \citep[p.~237, Fig.~6.16]{Pope:2000}, estimates of
$\langle (\partial v/\partial t)^2 \rangle$ from the Modane data should be accurate.
(Although Fig.~6.16 in Pope's book is drawn for the case
$R_\lambda =600$, a similar plot in the case $R_\lambda =2500$ is almost
indistinguishable from the case $R_\lambda =600$.)
Note that the value of $\eta$ given
in Table 1 of \citet{Malecot:2000} is in error, the correct value is given
in Yann Mal\'ecot's thesis and also in \citet{Kahalerras:1998} 
and in \citet{Gagne:2004} where the same Modane data is used.  
\medskip

In summary, it has been shown that the energy cascade rate $\eps = 1.8$ J/kg-s obtained 
by Kolmogorov's four-fifths law is in rough agreement with the energy dissipation rate
$\eps_{\rm diss} = 2.6$ J/kg-s obtained from equation (\ref{diss}).
\medskip

\section{Illustration using solar wind data}

\indent\indent 
In this section, we present two examples to illustrate the convergence of third-order 
moments for solar wind data.  The first 
example uses 1 second data for the radial magnetic field component $B_R$ measured
above the poles of the sun by the Ulysses spacecraft. 
The second example uses 64 second data for the radial
solar wind velocity $v_R$ measured in the ecliptic plane
near 1 AU by the ACE spacecraft.

\subsection{Analysis of $B_R$ using Ulysses data}

The radial magnetic field component  $B_R$ (in spacecraft RTN coordinates) was chosen 
because it enters the third-moment $\langle |\delta \bm B|^2\delta B_R \rangle$
that appears in the law for the cross-helicity cascade in MHD turbulence
\citep{Podesta_Forman:2007,Podesta:2008}. 
Ulysses data was chosen because it
is devoid of magnetic sector crossings which are usually  present
in data acquired near the ecliptic plane.  
The third-moment $\langle (\delta B_R)^3 \rangle$
changes algebraic sign in outward and inward magnetic sectors.  For this
reason, the presence of different magnetic sectors significantly complicates the 
analysis of this third-order moment.
\medskip

The Ulysses data selected for analysis consists of a time series of $\sim 1$ s data 
from the vector-helium magnetometer 
\citep{Balogh:1992} spanning the time interval from 1 July 1994 to
1 October 1994, 92 days. During this time
Ulysses distance from the sun decreased from 2.80 AU to 2.17 AU as
its heliographic latitude remained between $-70$ and $-81$ degrees.
The reversal of the solar magnetic field in the southern hemisphere was completed
in 1992 \citep{Snodgrass:2000} so the data used here contains only one
magnetic sector.  The time tags on the data show some data have a 1/2 second cadence, 
some data have a 1 or 2 second cadence, and there are also data gaps of various sizes.
The 1/2 sec data is downsampled to 1 sec and the data gaps are left intact
to create a time series with a uniform cadence of 1 sec.  Times when data
are missing are marked with fill values (such data are easily omitted from the analysis).
There are a total of $7.95\times 10^6$ data points in the time series and 23.55\%
of these points are missing data markers (fill values).  The average value of $B_R$
for the entire time series is $-0.526$ nT.
\medskip

The power spectrum for the Ulysses data shown in
Figure 5 is strikingly similar to that of the wind tunnel data in Figure 1.
\begin{figure}[t]
\vspace*{2mm}
\begin{center}
\includegraphics[width=8cm]{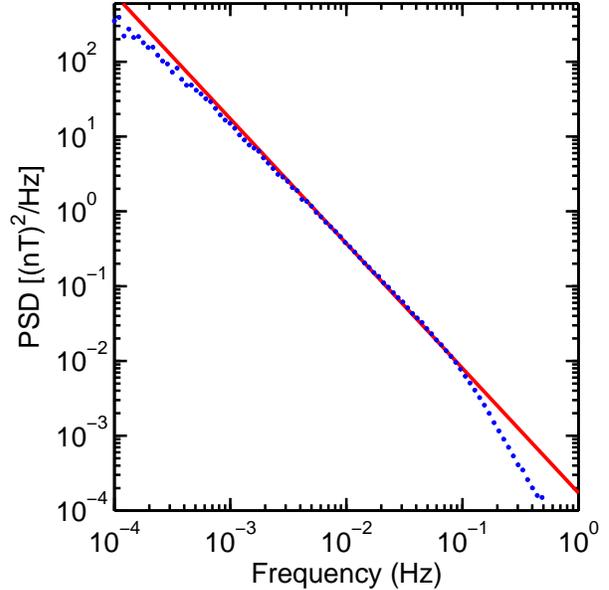}
\end{center}
\caption{Power spectrum of the radial magnetic field component $B_R$
for the Ulysses data used in this study (blue dots).  
The red line is proportional to $f^{-5/3}$.  The inertial range extends from less 
than $10^{-3}$ Hz to approximately $10^{-1}$ Hz.  The onset of the 
dissipation range is indicated by the change in slope around $10^{-1}$ Hz.}
\end{figure}
From Figure 5, the inertial range appears to extend from less 
than $10^{-3}$ Hz to approximately $10^{-1}$ Hz.  The time lag $\tau=60$ seconds
or, equivalently, $f=\tau^{-1}=1.67\times 10^{-2}$ Hz
is chosen for analysis because it lies inside the inertial range.
The same procedures used to analyze  the Modane data are employed for the
Ulysses data except that missing data is excluded from the analysis.
Consequently, a data block of size $M$ contains less than $M$ samples
(because of the presence of fill values) and
the actual number of samples $N$ varies from block to block.  Only those data
blocks where $N\ge 0.55 M$ are included in the analysis and the average number of samples
$N$ taken over all blocks of a given size $M$ is defined to be the sample size $N$
for that run.
\medskip

The results of the statistical analysis of Ulysses data for the time lag
$\tau=60$ seconds are shown in Figure 6.
\begin{figure}[t]
\vspace*{2mm}
\begin{center}
\includegraphics[width=6cm]{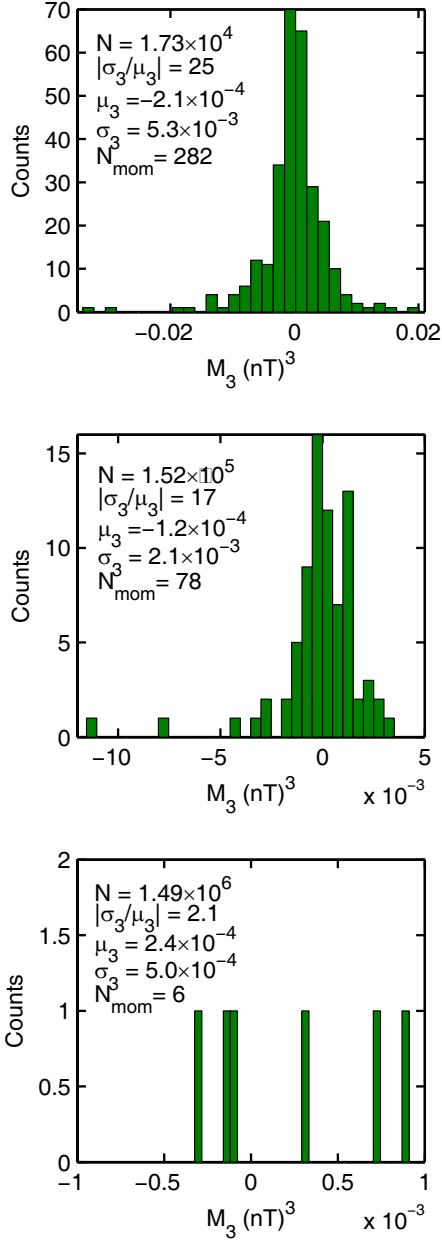}
\end{center}
\caption{Results from the analysis of the third order moment $\langle(\delta B_R)^3 \rangle$ 
for the time lag $\tau = 60$ s obtained using data from
Ulysses first southern polar pass.  The number $N$ is the approximate sample size used
to compute the third-moment and $N_{\rm mom}$ is the number of third-order moments
used to compute the statistics $\mu_3$ and $\sigma_3$.  The mean and standard 
deviation of $M_3(N)$, denoted by $\mu_3$ and $\sigma_3$, have units (nT)$^3$.  The value 
of $M_3$ obtained using the entire data record is $-8.9\times 10^{-5}$ (nT)$^3$.}
\end{figure}
The sizes of the data blocks used in the analysis are $M=2\times 10^4$, $2\times 10^5$, 
and $2\times 10^6$.  The offset from one data block to the next is $M$ for the upper 
plot and $M/2$ for the other two plots.  The algebraic sign of $\mu_3$ is negative
except in the lower plot, however, the sample size in the lower plot is too small to yield
adequate statistics.  The value of the third-moment
obtained using the entire data record is $-8.9\times 10^{-5}$ (nT)$^3$, a very small value. 
To gain some idea of the error, the error of the mean $\sigma_3/N_{\rm mom}^{1/2}$ 
estimated from Figure 6 is roughly $2\times 10^{-4}$  (nT)$^3$.
\medskip

The theoretical relation (15) may be evaluated using estimates obtained from the
data for the third-moment $M_3 = -8.9\times 10^{-5}$ (nT)$^3$, 
the sixth-moment $M_6 = 2.1\times 10^{-2}$ (nT)$^6$, and the correlation time
$\tau_c \simeq 21$ s determined from the autocorrelation function for the
sequence $(\delta B_R)^3$.  The values of the moments are uncertain, especially
higher order moments such as $M_6$ which can be strongly affected by the presence
of outliers in the data \citep{Horbury_Balogh:1997}, however, they are used anyway
to explore the fit to the data of the relation (15).  Thus, the theoretical
relation (15) takes the form
\begin{equation}
\left| \frac{\sigma_3(N)}{\mu_3(N)}\right| \simeq \frac{7500}{\sqrt N}.
\label{fit2}
\end{equation}
For the three runs shown in Figure 6 the theoretical values for the ratio
$|\sigma_3/\mu_3|$ are 57, 19, and 6.  These are roughly consistent with the values
found in Figure 6.  For all data, $N\simeq 6\times 10^6$ and
the relation (\ref{fit2}) predicts a relative errror $|\sigma_3/\mu_3|\simeq 3$.
Hence, $\mu_3 \simeq -8.9\times 10^{-5} \pm 2.7\times 10^{-4}$ (nT)$^3$, which
is consistent with the estimates in the preceeding paragraph.
\medskip

In summary, the large magnitude of the ratio $\sigma_3/\mu_3$  
is partly due to the fact that $\mu_3$ is close to zero and this makes 
it impossible to obtain adequate convergence with the limited data used in this study.  
One may conclude from these results that
a much larger data set than the one used here is needed to determine the
third-moment of $\delta B_R$ accurately.   Nevertheless, the results presented here are
still useful for determining approximate upper and lower bounds for this
third-order moment.

\subsection{Analysis of $v_R$ using ACE data}

Solar wind measurements of the radial velocity component from the Advanced Composition 
Explorer (ACE) are analyzed in the same way.  The ACE spacecraft is in orbit 
around the Sun-Earth $L_1$ libration point $240R_e$ sunward of the Earth.
The ACE SWEPAM plasma instrument has a 64-second cadence \citep{McComas:1998} and we use all 
data available during the three year period from 2005 through 2007, about 1.4 million 
data points.  Note that solar minimum is expected to occur in late 2008 or early 2009.  
Non-overlapping data blocks of 100 points (about 10,000 blocks) to 256,000 data points 
(5 blocks) are used.  Each data block may include fill values
(missing data markers) that are present in the time series.  All fill values are omitted
from the analysis and any data block in which the number
of fill values exceeds 10\% is excluded from analysis.  Third-order moments of 
$\delta v_R$ are calculated for two different time lags, 
$\tau =256$ seconds and $\tau= 2048$ seconds.  The inertial range 
in the ecliptic plane near 1 AU extends from about 1 second to about 1 hour and, therefore, both
of these time lags lie in the inertial range.
\medskip

Figure 7 
\begin{figure}[t]
\vspace*{2mm}
\begin{center}
\includegraphics[width=6.6cm]{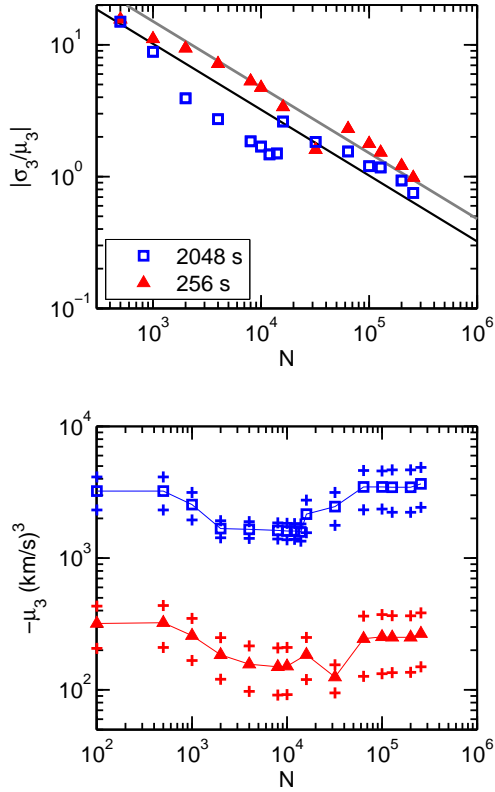}
\end{center}
\caption{Results from the analysis of the third-order moment $\langle(\delta v_R)^3 \rangle$ 
using ACE data for 2005--2007. Red triangles and blue squares correspond to the time lags
$\tau=256$ s and $\tau=2048$ s, respectively. The upper plot shows the convergence of the ratio
$|\sigma_3/\mu_3|$ as a function of the number of samples $N$; the solid lines are the
theoretical predictions from equation (15), $477/N^{1/2}$ and $322/N^{1/2}$.  
The lower plot shows the values of the third-order moment plus and minus the
error in the mean, that is, $\mu_3 \pm \sigma_3/N_{\rm mom}^{1/2}$, where
$N_{\rm mom}$ is the number of moments used to compute the mean.}
\end{figure}
shows that in the solar wind, the ratio of the standard deviation of the third-moment 
to the average value of the third-moment $|\sigma_3/\mu_3|$ has the same $N^{-1/2}$ 
dependence predicted by equation (15) as does the wind tunnel data analyzed in Section 5. 
Remarkably, the amplitude of this relation is quantitatively similar for both
wind tunnel data and solar wind data, even though the solar wind has a fast/slow stream structure
and the turbulence is magnetohydrodynamic in nature.
As with the wind tunnel data, it appears from Figure 7 that more than $10^7$ solar wind velocity
measurements are needed to accurately determine the third-order moment 
$\langle (\delta v_R)^3\rangle$ (error less than 10\%).
However, sample sizes $N\sim 10^6$ may give sufficient accuracy for some
applications.
\medskip

Also plotted in the upper plot in figure 7 are the theoretical curves, equation (15), for the 
two time lags studied.  A rough estimate of equation (15) is obtained by using all data in the
record to estimate the sixth-moment of $\delta v_R(\tau)$, $M_6$, the third-moment $M_3$, and the 
correlation time $\tau_c$ of the sequence $[\delta v_R(\tau)]^3$.  For the time lag $\tau=256$ s, 
this yields $M_6\simeq 8.8\times 10^9$ (km/s)$^6$, $M_3\simeq 2.5\times 10^2$ (km/s)$^3$, 
and $\tau_c/\tau \simeq 0.4$.  In this case, $n=1.6$ and equation (15) becomes
\begin{equation}
\left| \frac{\sigma_3(N)}{\mu_3(N)}\right| \simeq \frac{477}{\sqrt N}.
\end{equation}
For the time lag $\tau=2048$ s, 
$M_6\simeq 9.5\times 10^{10}$ (km/s)$^6$, $M_3\simeq 3.4\times 10^3$ (km/s)$^3$, 
and $\tau_c/\tau \simeq 0.4$.  In this case, $n=12.8$ and equation (15) becomes 
\begin{equation}
\left| \frac{\sigma_3(N)}{\mu_3(N)}\right| \simeq \frac{322}{\sqrt N}.
\end{equation}
These represent reasonable asymptotic fits to the data shown in Figure 7 as $N$ becomes large.
\medskip

What accuracy can be claimed for measurements of this third-order moment in the solar wind?  
The upper plot in Figure 7 indicates that for $N\simeq 1.4\times 10^6$ data points
the error is around $\sim 40\%$.  The lower plot in
Figure 7 shows the mean values of the third moment $\mu_3$ plus and minus the 
standard error of the mean $\sigma_3/N_{\rm mom}^{1/2}$, where $N_{\rm mom}$ is the 
number of moments used to compute the mean.  It appears that the relative 
error is roughly the 
same at both lags and that within the calculated errors the third moment is 
proportional to lag.  Such proportionality is most clearly demonstrated
by computing the third-moment as a function of the time lag $\tau$ (not shown).
\medskip

It is interesting that both $R_\lambda \simeq 2500$ wind tunnel data and
solar wind velocity data require roughly the same number of data points to
obtain good convergence of third-order moments for time lags in
the inertial range.  In part, this may be because both
kinds of turbulence have similar Reynolds numbers.  The Reynolds number
in the solar wind can crudely be estimated using the hydrodynamic relation
$L/\eta=\Rey^{3/4}$, where $L$ is the integral scale, $\eta$ is the Kolmogorv scale
(dissipation scale), and $\Rey$ is the Reynolds number based on the integral scale.
Solar wind power spectra indicate that $L/\eta\sim 10^5$ and, therefore the 
Reynolds number is of order $10^6$.  This is equivalent to a Taylor-scale
Reynolds number 
\begin{equation}
R_\lambda = \big( {\textstyle \frac{20}{3}} \Rey\big)^{1/2} \sim 2600
\end{equation}
\citep[p.\ 200, eqn.\ 6.64]{Pope:2000}.
Thus, the Reynolds numbers of Modane wind tunnel data and solar wind velocity data
at 1 AU are similar. 

%
%
\medskip


\section{Conclusions}

The purpose of this study is not to compute turbulent energy cascade rates using
third-order moments.  
The purpose of this study is to show how the accuracy of third-order moments
can be estimated and controlled.  A simple theory is presented that describes the statistical
convergence of third-order moments, such as $\langle (\delta v_\parallel)^3\rangle$,
as a function of the record length.  An important conclusion is that the accuracy of 
third-order moments depends on the number of correlation lengths spanned by the 
time series as expressed by equations (14) and (15).  The techniques described here are
useful for assessing the accuracy of third-order moments obtained using measured data.
Practical applications of the theory
have been illustrated using wind tunnel data and solar wind data.  
\medskip

For the accurate calculation of third-order moments from wind tunnel data 
with a Taylor-scale
Reynolds number $R_\lambda \simeq 2500$, the number of data points required to obtain
an error less than 10\% at the time lag $\tau=20$ ms is $N=10^7$ or, equivalently,
a record length spanning $N_c=10^5$ correlation lengths.
For fluctuations of the radial solar wind velocity $v_R$, the analysis 
of ACE data in the ecliptic plane near 1 AU shows that for the time lags $\tau=256$ s
and $\tau=2048$ s the number of data points required for an accurate 
determination of the third-order moment is also roughly $N=10^7$. This is equivalent to
$N_c=6.3\times 10^6$ and $7.8\times 10^5$ correlation lengths for $\tau=256$ s
and $2048$ s, respectively.
However, $N\sim 10^6$ data points may yield sufficient accuracy for some applications.  
\medskip

For fluctuations of the radial magnetic
field component $B_R$ over the poles of the sun at a heliocentric distance of approximately
2.5 AU, the value of the third-order moment is close enough to zero that convergence
of the third moment could not be demonstrated using an interval of Ulysses data
with approximately six million points (not including fill values), a record consisting
of approximately $3\times 10^5$ correlation lengths. 
This suggests that third-order moments of solar wind magnetic field components
must be computed carefully because 
without a sufficiently large number of data points and without evaluation of the probable errors 
using eq (15) the calculation of the third-order moments are not meaningful.
\medskip

It should be noted that the above two examples based on solar wind data from Ulysses and ACE are 
distinctly different from each other and from the example based on Modane wind tunnel data.  
The Ulysses study pertains to the radial magnetic field component at high heliographic latitudes 
and the ACE study pertains to the radial velocity component in the ecliptic plane.   
These studies provide two separate examples of the estimation of third-order moments
and their uncertainties using solar wind data.    
For wind tunnel data, Kolmogorov's four-fifths law predicts the third-order velocity moment scales linearly
in the inertial range.  For solar wind data, neither the third-order moment of the radial velocity 
component nor the third-order moment of the radial magnetic field component is
predicted to scale linearly.  The latter examples simply serve to illustrate the application of
statistical convergence techniques to solar wind data.  It is of interest to note, however,
that the third-order moment of the radial velocity component in the solar wind was found to 
scale approximately linearly in the study by MacBride et al. (2008) and appears to provide the dominant
contribution to the energy cascade rate estimated from the scaling relations of Politano and 
Pouquet (1998).
\medskip

In conclusion, the present study has some noteworthy implications for measurements 
of the energy cascade rate in the solar wind.
Empirical estimates of the energy cascade rate in the solar wind have recently been obtained 
under the assumption that the turbulence is approximately incompressible and isotropic 
so that the third-moment scaling relations of Politano and Pouquet (1998) for homogeneous isotropic
incompressible MHD turbulence could be applied.  In the studies by 
\citet{Sorriso-Valvo:2007} and \citet{Marino:2008} the number of data
points employed to compute the required third-order moments was around
2000.  As shown in the present study, this number is insufficient to obtain accurate
estimates of third-order moments in the solar wind.  This may explain
why \citet{Sorriso-Valvo:2007} and \citet{Marino:2008} did not find a linear scaling of
the third-order moments in some of the intervals they studied, 
and why they found it was rare for
linear scaling to be observed simultaneously for both of the Elsasser variables.
Although MacBride et al. (2008) did not use the convergence tests proposed here,
they used large enough data sets that the third order moments in the
Politano and Pouquet scaling laws became insensitive to adding more data.
To obtain stable estimates of the third-moments, this convergence criterion required the use 
of at least one year of ACE plasma and magnetic field data, roughly $10^6$ data points.
In the future, the convergence of third-order moments and the associated error
estimates that such convergence studies provide should become an integral
part of any analysis of solar wind data involving third-order moments.



\begin{acknowledgements}
We are grateful to Bernard Vasquez for helpful discussions.  
Support for this work was provided by NASA grant NNX08AJ19G under the Heliophysics
Guest Investigator Program. Charles Smith acknowledges NASA support from the ACE program.
We thank the ACE Science Center, the ACE/SWEPAM team, and the Space Physics Data Facility at
Goddard Space Flight Center for providing the solar wind data used in this study.
\end{acknowledgements}

\bibliographystyle{copernicus}
\bibliography{jp}

\IfFileExists{\jobname.bbl}{}
 {\typeout{}
  \typeout{******************************************}
  \typeout{** Please run "bibtex \jobname" to optain}
  \typeout{** the bibliography and then re-run LaTeX}
  \typeout{** twice to fix the references!}
  \typeout{******************************************}
  \typeout{}
 }


\end{document}